\documentclass[12pt,onecolumn,showpacs,amssymb,aps,nofootinbib,floatfix]{revtex4-1}
 \usepackage{epsfig}
 \usepackage{xcolor}
 \newcommand{\eqref}[1]{Eq. \ref{#1} }
 \newcommand{\figref}[1]{Fig. \ref{#1} }
 \newcommand{\secref}[1]{section \ref{#1} }
   \newcommand{\appref}[1]{Appendix \ref{#1} }

\newcommand{\eqcomma}{\phantom{AA},\phantom{AA}}
\newcommand{\bra}[1]{  \left| #1 \right> }
\newcommand{\ket}[1]{  \left< #1 \right| }

\begin{document}
\title{Spin alignment of vector mesons as a probe of spin hydrodynamics and
    freeze-out}
\author{Kayman J. Gon\c{c}alves,Giorgio Torrieri}
\affiliation{Universidade Estadual de Campinas - Instituto de Fisica "Gleb Wataghin"\\
Rua Sérgio Buarque de Holanda, 777\\
 CEP 13083-859 - Campinas SP
}
\begin{abstract}
  We argue that a detailed analysis of the spin alignment of vector mesons can serve as a probe of two little-understood aspects of spin dynamics in the vortical fluid:  The degree of relaxation between vorticity and particle spin polarization, and the degree of coherence of the hadron wavefunction at freeze-out.
  We illustrate these with a coalescence model.
  \end{abstract}
\maketitle
\section{Introduction\label{intro}}
\color{black}
The experimental detection of event plane polarization in heavy ion collisions has generated a great amount of experimental and theoretical interest \cite{lisa}.   The data on global polarization is reasonably described by an ideal hydrodynamic model where angular momentum is transferred to spin at freeze-out following an extension of the Cooper-Frye equilibrium formula \cite{cf}.    While experimentally this is a great success (although the {\em local} rather than {\em global} polarization is less well described \cite{lisa}), from a theory point of view such a model cannot be the whole story:
It is now a commonly accepted idea  \cite{uscausality,uscas2,relax,jeon,noronha,analspin1,analspin1.5,analspin2}
that spin and vorticity are not always in equilibrium or aligned (in non-relativistic physics this relaxation is known as the Fert-Valet equation \cite{fart,fart2}).
There is however little phenomenology associated with this lack of alignment in the context of $\Lambda$ polarization.   Such a phenomenology is complicated by the fact that spin and angular momentum are not separately conserved.  Hence, the detail of how angular momentum is partitioned at freeze-out can not be straight-forwardly be linked by a Cooper-Frye type ansatz, that depends on entropy conservation and conservation laws only \cite{cf}.   In principle more complicated dynamics (such as coalescence \cite{coal1,coal2,coal3}), even if close to global equilibrium, will imprint a non-trivial signature.

In this work we would like to propose a detailed measurement of the spin space density matrix of vector meson, projected on the global event coordinate system (with $\theta$ defined by the reaction plane and $\phi$ on beam axis).\color{black}
The experimental measurement of spin alignment of vector mesons ($K^*$ and $\phi$) along the reaction plane in heavy ion collisions is a nice complement to the measurement of hyperon polarization \cite{alice,star}.   However, reconciling them in a single framework is still an open problem, both from the theoretical and the experimental point of view\cite{lisa}.

Experimentally, spin alignment and polarization are not really the same observable, although they both depend on analyses of angular distributions of decay products in the mother-particle's rest frame.  Polarization measurements are done using  weak decays of hyperons, and, because of parity violation in such decays, the experimentalist can get a direct handle on the spin direction.   Spin alignment on the other hand is done using parity-conserving strong $p-$wave decays.  As such, as the name suggests, it is impossible to distinguish ``up pointing'' from ``down-pointing decays'', but just distinguish up/down third quantum number $m_s \pm 1$ (aligned) vs $m_s=0$ spin Eigenstates.

However, this is not to say that spin alignment potentially provides less information than polarization.    p-wave decays do not just have one reference axis but two:  In the rest frame of the decaying meson \cite{seyboth,LI} the angle $\theta$ is defined as the angle between the polarization vector and the momentum of the daughter particle (either of them, since spin alignment will be the same for $\theta,\pi-\theta$).   There is however an additional azimuthal angle $\phi$ which, provided an event-by-event reference can be provided, can be used to get information about the decaying meson and the underlying event.
 in the reference frame at rest with the meson decay, given a definition for an axial angle $\theta$ and azimuthal angle $\phi$ the distribution of decay products in terms of $\theta$ and $\phi$
\cite{seyboth,LI}, presented as a polynomial of harmonic coefficients, is:
\begin{equation}
W(\theta,\phi) \propto
\label{dis}  \cos^{2}\theta\rho_{00}+\sin^{2}\theta \left(  \frac{1-\rho_{00}}{2}+  r_{1,-1} \cos(2\phi)+\alpha_{1,-1} \sin(2\phi) \right)+ \sin(2 \theta) \left( r_{10} \cos\phi +\alpha_{10} \sin\phi\right)
 \end{equation}
where the coefficients to the trigonometric formulae are related to the density matrix elements of the meson ensemble via a normalization coefficient $3/(4\pi)$ and
\begin{equation}
  \begin{array}{ccc}
\mathrm{Variable} &    \mathrm{Element}& \mathrm{coefficient}\times \frac{3}{4\pi}\\
 \rho_{00} &    \rho_{00} & \cos^2\theta\\
 \frac{1-\rho_{00}}{2} &   \frac{\rho_{11}+\rho_{-1-1}}{2} & \sin^2 \theta\\
    r_{10} & Re[\rho_{-10}-\rho_{10}] & \sin(2\theta)\cos(\phi)\\
   \alpha_{10} &   Im[-\rho_{-10}+\rho_{10}] & \sin(2\theta)\sin(\phi)\\
   r_{1,-1} &     Re[\rho_{1,-1}] & \sin^2 \theta \cos (2\phi)\\
   \alpha_{1,-1} &     Im[\rho_{1,-1}] & \sin^2 \theta \sin(2\phi)
  \end{array}
  \label{tablecoeff}
\end{equation}
It is clear that for a mixture of up/down and zero states w.r.t. a single direction and this direction is known (the reaction plane) $\phi$ is irrelevant because the density matrix will be in a diagonal form and there will be no $\phi$ dependence unless the ``correct'' spin axis is mis-identified (\cite{xia} points out this is indeed the case if both longitudinal and transverse polarization occurs and the reaction plane is used an axis).
If, however, more than two sources of angular momentum (vorticity and spin currents!) somehow combine to produce vector meson polarization, $\phi$ distribution could be non-trivial in any axis because the density matrix reconstructed from \eqref{tablecoeff} is not pure.

\begin{figure}[t]
        \begin{center}
     \epsfig{width=0.89\textwidth,figure=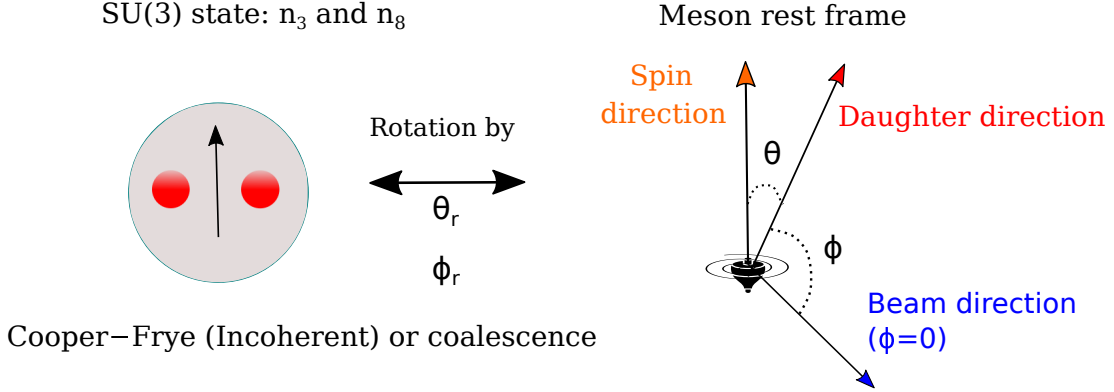}
          \caption{\label{alignment} Relation between the angles $\theta,\phi$ as measured in experiment and the meson's density matrix parametrized as in \eqref{sigma3def} }
        \end{center}
\end{figure}

So far, experimental collaborations \cite{alice,star} only measured the $\theta$ angle distribution w.r.t. the reaction plane (with the $\rho_{00}$ alignment factor), since this measurement can be linked to global vorticity.    $\rho_{00}$ can then be used as a related parameter to the polarization probability.
One event-independent easy to implement definition for the other angle would be in terms of the beam axis (\figref{alignment}), although others have bee used in Charmonium studies in p-p collisions \cite{jpsi}.  However, although extra analysis has been suggested by theorists \cite{oliva,phicoal,xia}, no systematic interpretation of the extra parameters has so far been proposed.
In this work we attempt to do this, in terms of simple popular models of hadron production, such as coalescence \cite{coal1,coal2,coal3} and Cooper-Frye (\cite{cf} and references therein), as well as some general properties of spin and angular momentum. 

The spin state of a particle of spin $N=2s+1$, projected on spin space, will be an $N\times N$ density matrix, described as \cite{gendens}
\begin{equation} 
\rho = U \Lambda U^{-1} \eqcomma U \in U(2s+1) \eqcomma \lambda=diag[\lambda_i] \eqcomma \sum_{1}^{2s+1} \lambda_i=1
\end{equation}
$N$ here generically corresponds to the internal state space, but of course it reduces to rotations between spin Eigenstates when the density matrix considered is one projected in spin space.

Assuming we have a measured distribution of spin observables we could attempt to reconstruct such a density matrix.
The amount of information that can be extracted from it ( to what extent it is coherent w.r.t. any particular direction, the parameter space constraints) will depend on "Coadjoint orbits", coset spaces w.r.t. the 
subgroups of $U(N)$ describing rotations between degenerate Eigenstates.   Rotations within the same orbit keep the density matrix diagonal.

For a spin 1/2 particle $\hat{\rho}_2$ the density matrix reduces to
 one ``qubit'' of information, a point on the the Bloch sphere.
One crucial issue, specific to qubits, is that the Bloch sphere describes the parameter space of both pure and mixed qubits.  This is however a specific fact of qubits \cite{preskill}, that there always exists a unit 3-vector $\hat{n}$ such that 
\begin{equation}
  \label{qubit}
  \hat{\rho}_2 = \frac{1}{2}\left( I+ \hat{n}.\vec{\sigma}\right)
\end{equation}
where $\hat{\sigma}$ are the Pauli matrices.  
No matter what dynamics is used to combine spin (call it $\hat{n}_1$) and angular momentum (call it $\hat{n}_2$) from the quark-gluon plasma into the hadron spin, the resulting density distribution for 
$\hat{\rho}_2$ will be a pure density state for some axis (defined by $\hat{n}$, perhaps a mixture of $\hat{n}_1$ and $\hat{n}_2$ as in \cite{xia}).    A rotation to this axis will purify the density matrix.
Hence, how spin and angular momentum of quarks and gluons combine to produce the $\Lambda$ spin 
(is it a unitary process or interaction with an environment?) is in principle inaccessible.
Of course the dependence of that qubit on that particle's momentum and rapidity might tell us
something in this direction but this necessitates fitting and will always be model-dependent in a non-straight-forward way.

These considerations are however specific to spin 1/2 fermions with two Eigenstates\footnote{In a physically very different but mathematically similar context, the explanation as to why two flavors are not enough for CP violation in the Cabibbo-Kobayashi-Maskawa matrix is similar}.
The density matrix of a spin 1 particle, in contrast is (left side of \figref{alignment}) projected to internal spin space is a qutrit \cite{qutrit1,qutrit}, where pure and mixed states do not have this degeneracy and coherence can be evaluated by geometrical means.   
It is well known that \cite{qutrit1} that a qutrit`s density matrix can be parametrized as
\begin{equation}
  \label{sigma3def}
\rho = \frac{1}{3} \left( 1+\sqrt{3} \vec{n}_8.\vec{\lambda} \right)
  \end{equation}
where $\vec{n}_8$ is an 8-dimensional vector (not representable in configuration space)  and $\vec{\lambda}$ represents the Gell-Mann density matrices.
This is a parametrization, rather than a representation, since 
the coefficients do not in any way transform under $SU(3)$ algebra and spin operators transform as
 SU(2) representations, but this parametrization can be used to 
characterize the density matrix, specifically its degree of coherence and alignment.

\begin{figure}[h]
	\begin{center}
		\includegraphics[width=0.50\linewidth]{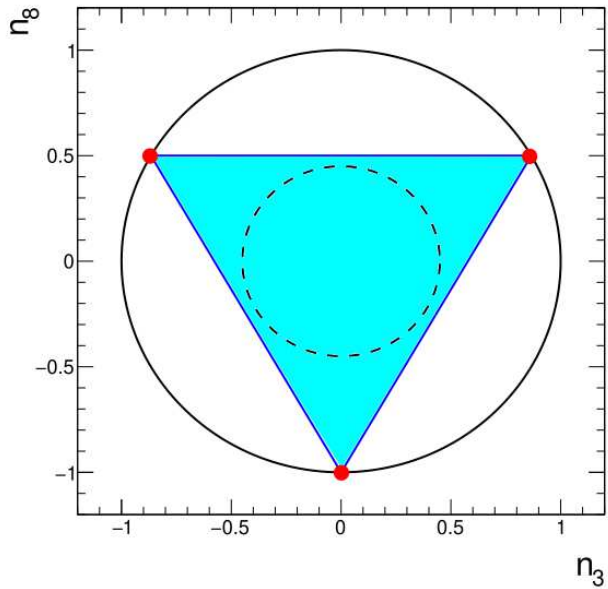}
		\caption{(color online) A representation of the states of the density matrix Eq. \ref{rho8}. The points touching the outer circle correspond to pure states, and the points inside the blue triangle are the mixed states, for example, the dotted circle points inner the triangle.}
		\label{newstates}
	\end{center}
\end{figure}

Aligned in the $8$ direction ($\vec{n}_8$ has only the 8th component), this matrix is going to look like
\textcolor{black}{
\begin{equation}
  \label{rho8}
  \rho_8(n_3,n_8)=\frac{1}{3}\left( \begin{array}{ccc}
  1+\sqrt{3}\;n_3+n_8 & 0 & 0 \\
  0 & 1-\sqrt{3}\;n_3+n_8 & 0\\
  0 & 0 & 1-2n_{8}
  \end{array} \right)
\end{equation}
}
this contains two qubits (more exactly, a ``qutrit'') of information \cite{qutrit,geom}.
The additional purity constraint  $\mathrm{Tr} \left(\hat{\rho}^2\right)=\hat{\rho}$ gives
 \begin{equation}
   \label{det}
\left(1+n_8+\sqrt{3}\;n_3\right)\left(1+n_8-\sqrt{3}\;n_3\right)\left(1-2\;n_8\right)=0
\end{equation}
 Thereby, we will have the three points $\left(n_3,n_8\right)=(0,-1),\;\frac{1}{2}(\pm\sqrt{3},1)$ that represent the pure states.
 In the figure \figref{newstates}, we show that the  points where the triangle touches the outer circle.
 Conversely, the points on the inner circle of \figref{newstates} correspond to maximum
 entropy
 \begin{equation}
   \label{maxmixed}
\hat{\rho} = \frac{\mathcal{F}}{\mathcal{F}^2+\mathcal{F}+1}\mathrm{Diag} \left[\mathcal{F},1,\frac{1}{\mathcal{F}} \right] \eqcomma \mathcal{F} \sim \exp\left[\frac{\Omega}{T}  \right]
   \end{equation}
 expected from the Cooper Frye type freezeout.  Note that in the frame co-moving with the flow the density matrix examined in \cite{xia} has this form.

If one can
\begin{description}
\item[associate] each qutrit to an axis direction, with both $\theta$ and $\phi$ angles according to \eqref{dis}
\item[understand] how plasma spin and orbital angular momentum combine into hadronic spin in a calculable way
\end{description}
one could potentially come up with a phenomenology of spin-orbit misalignment from a fit to Eq. \ref{dis}.

The association is in fact straight-forward:
It is known experimentally from $\Lambda$ polarization \cite{lisa} that
spin polarization has both a transverse and a longitudinal component.
Physically, their origins are very different:
The transverse component is the result of the initial impact parameter, while
the longitudinal component is the result of the same hydrodynamic evolution that generates azimuthally anisotropic flow coefficients.

This means that if there is diffusion between vorticity and spin, the admixture of spin vs
angular momentum in the two components is generally very different.
Naively for the transverse component vorticity and spin will be in
equilibrium, the longitudinal component less so (the opposite of flow,
where longitudinal flow is formed initially, transverse flow is not).
Considering that this difference in equilibration is directly correlated to axes, it is worth seeing if the extra information in vector mesons is obtainable from an angular analysis including these axes.

Now, the second requirement, understanding how plasma spin and angular momentum combine into vector meson spin, is a bit more complicated and model dependent.
The rest of the paper is about this topic.      In section \ref{spinexp}, we relate experimental coefficients to group-theoretical parameters.
In section \ref{spincoal} , we relate these parameters to the spin density and vorticity within a coalescence type picture.

\section{From qutrits to experimental angular distributions \label{spinexp}}
Consider a spin 1 density matrix aligned in a certain direction.  We need to understand the relation between a physically natural reference frame, e.g. Eq. \ref{rho8}, to the laboratory coordinate system.   The two are related by  an additional transformation parametrized by angles $\theta_r,\phi_r$ (\figref{alignment})
\begin{equation}
\label{rhogen}
  \rho(n_3,n_8,\theta_r,\phi_r)= U(\theta_r,\phi_r) \rho_8(n_3,n_8)  U(\theta_r,\phi_r)^{-1}
\end{equation}
where\footnote{Note the use of Clebsch-Gordon coefficients in this and the next section, to underline that while one can parameterize the density Matrix using Gell-Mann matrices, it still transforms as a representation of SU(2).}
\textcolor{black}{
	\begin{equation}
		U(\theta_r,\phi_r)=D\left(\alpha,\beta,\gamma\right) = exp\left(\frac{-i J_{z} \alpha}{\hbar}\right)exp\left(\frac{-i J_{y} \beta}{\hbar}\right)exp\left(\frac{-i J_{z} \gamma}{\hbar}\right)
	\end{equation}
}
\textcolor{black}{
  Making $\alpha = \phi_r$, $\beta = \theta_r$ and $\gamma = 0$ we can decompose
  \begin{equation}
    \label{dmat}
    D^{j}_{m^{'} m }(\phi_r,\theta_r) = \ket{j,m^{'}}D\left(\phi_r,\theta_r,0\right)\bra{j,m}
\end{equation}
    with
    \begin{equation}
      \label{dexp}
		D^{j}_{m^{'} m }(\phi_r,\theta_r)= e^{-i m^{'}\phi_r}d^{j}_{m^{'} m }(\theta_r)\eqcomma \left[D^{j}_{m^{'} m }(\phi_r,\theta_r)\right]^{-1}= e^{i m\phi_r}\left[d^{j}_{m^{'} m }(\theta_r)\right]^{-1}
	\end{equation}
}
and for $j=1$, we have explicitly	
\textcolor{black}{
	\begin{equation}
		\label{d}
		d^{1}(\theta_r)=\left( \begin{array}{ccc}
		\frac{1}{2}(1+\cos \theta_r) & -\frac{1}{\sqrt{2}}\sin \theta_r & \frac{1}{2}(1-\cos \theta_r) \\
		\frac{1}{\sqrt{2}}\sin \theta_r & \cos \theta_r & -\frac{1}{\sqrt{2}}\sin\theta_r\\
		\frac{1}{2}(1-\cos \theta_r) & \frac{1}{\sqrt{2}}\sin \theta_r & \frac{1}{2}(1+\cos \theta_r)
		\end{array} \right)
	\end{equation}
}

\textcolor{black}{Thereby, the right hand side of  \eqref{rhogen} becomes }
  \begin{equation}
    \label{rhogen2}
    =\sum_{m'' , m} \ket{j,m'}D\bra{j,m''}\ket{j,m''}\rho_8\bra{j,m}\ket{j,m}D^{-1}\bra{j,m'''}
  \end{equation}  
\[\
        =\sum_{m'' , m} e^{i(m'''-m')\phi_r}d^{j}_{m^{'} m^{''} }(\theta_r)\ket{j,m''}\rho_8\bra{j,m}\left[d^{j}_{m^{} m^{'''} }(\theta_r)\right]^{-1}
= \sum_{m'' , m}e^{i(m'''-m')\phi_r}T_{m',m'''}(\theta_r,n_3,n_8)
\]
Now, as discussed in the introduction given two event-invariant axes, experimentalists measure (right panel of \figref{alignment}) all the coefficients of \eqref{tablecoeff} via a Fourier analysis of \eqref{dis}.  
Comparing \eqref{tablecoeff} with \eqref{rhogen} we get a system of complicated-looking but actually solvable equations (although in practice
systematic effects might be difficult to disentangle \cite{practical}).
\textcolor{black}{ We can take advantage of the fact that $\hat{\rho}$ in Eq. \ref{sigma3def} is parametrized by Gell-Mann matrices, by choosing appropriate two-sections and three-sections . The sections represent the choice of the coefficients of the density matrix that we will consider, but your choice depends on the geometries possibles on the Block sphere generalization of a qutrit state \cite{qutrit}.
}

Choosing the two-sections we obtain, after some Algebra\footnote{  The operations leading to \eqref{sys1},\eqref{sys2} and \eqref{sys3} are not $SU(2)$ transformations} 
\begin{eqnarray}
\frac{1}{12} \left(3 \left(n_8-\sqrt{3}\;
n_3\right) \cos \left(2 \theta _r\right)-\sqrt{3}\;
n_3+n_8+4\right)=\rho_{00} \label{sys1}\\
\frac{\left(n_8-\sqrt{3} \;n_3\right) \sin
	\left(\theta _r\right) \cos \left(\theta _r\right) \cos
	\left(\phi _r\right)}{\sqrt{2}}=r_{10} \label{sys2}\\
-\frac{\left(\sqrt{3}\;n_3+3 n_8\right) \sin
	\left(\theta _r\right) \sin \left(\phi _r\right)}{3
  \sqrt{2}}=\alpha_{10} \label{sys3}
\end{eqnarray}
\begin{equation}
\label{phirdef}
  \phi_r=-\frac{1}{2}\tan^{-1}\left(\frac{\alpha_{1,-1}}{r_{1,-1}}\right)
\end{equation}
Which allows us obtain  the elements of the array in \eqref{tablecoeff}

\eqref{phirdef} is a trivial combination of observables.  The rest,
Solving the systems equations \ref{sys1} to \ref{sys3} we obtain the expressions for $n_{3,8},\theta_r$ as rational functions given in \appref{seclambda}.

It is a good point to pause and understand the physical meaning of these formulae.   The dynamics of transfer of angular momentum from collective vorticity to spin is complicated and still not understood ( \secref{spincoal} will give a simple example).  The relativistic regime complicates this considerably, since angles are not Lorentz invariant and hence spin directions potentially change when one moves from the co-moving frame with the resonance (where spin is imprinted) to the lab frame (where it is measured).   In this work, we ``evade'' such difficulties by only considering the density matrix in spin space, with momentum projected out.   The vorticity dynamics, {\em whatever it is}, will imprint itself into the density matrix elements, measurable experimentally through Eq. \ref{dis}, from which the expectation value (over all momentum space in the detector acceptance) of $\theta_r$ and $\phi_r$ will be obtained via \eqref{thetardef} and \eqref{phirdef}.   While of course for non-linear processes the average of a function is not a function of the average, we can reasonably hope that reconstructing information about both hydrodynamics and coherence is possible, i.e., $\theta_r$ and $\phi_r$ might not be wholly washed by event-by-event fluctuations.
The amount of such washing-out also depends on the coherence of the freeze-out process.
For instance, for a totally incoherent freeze-out,
Since the only axis that enters the Cooper-Frye formula is the vorticity axis, one expects $\phi_r$ to be washed out if freeze-out happens in this approach (this can be seen straight-forwardly from \eqref{phirdef}, since the ratio of the coefficients depends on $\phi_r$ only) and $\theta_r$ will have a peak consistent with zero (up to a shift by $\pi$ of course).  However in more complicated freeze-out dynamics this does not have to be the case.  A similar analysis is impossible for a Fermion density matrix projected on this system of coordinates as per the discussion around \eqref{qubit}.  This is a good illustration of the potential in looking at the full vector meson decay angular distribution in the global coordinate system.

Physically, $\theta_r,\phi_r$ are related to the transverse and longitudinal axis.  The first represents ``initial state'' vorticity, the second is created concurrently with longitudinal vorticity, during the hydrodynamic evolution stage \cite{lisa}.   A relative non-equilibrium therefore will manifest itself by comparing the $\rho_{00}$ and $\rho_{0i},\rho_{ij}$.   We therefore expect $\theta_r$ to be indeed peaked around the reaction plane $\theta_r \simeq 0$.   Assuming non-equilibrium between spin and angular momentum at freeze-out, $\phi_r$ will also exhibit a peak at some value {\em different} from zero (again, a distribution symmetric by $\pi$), set in detail by the non-equilibrium spin-vorticity diffusion. 

While a quantitative comparison is outside the scope of this work, in the next \secref{spincoal} we will demonstrate how
``early vorticity'' (expected to be mainly transverse) combines with later vorticity (which is also longitudinal) in a coalescence model.
\color{black}.
\section{From plasma spin and vorticity to qutrits via coalescence \label{spincoal}}
\begin{figure}[t]
        \begin{center}
     \epsfig{width=0.85\textwidth,figure=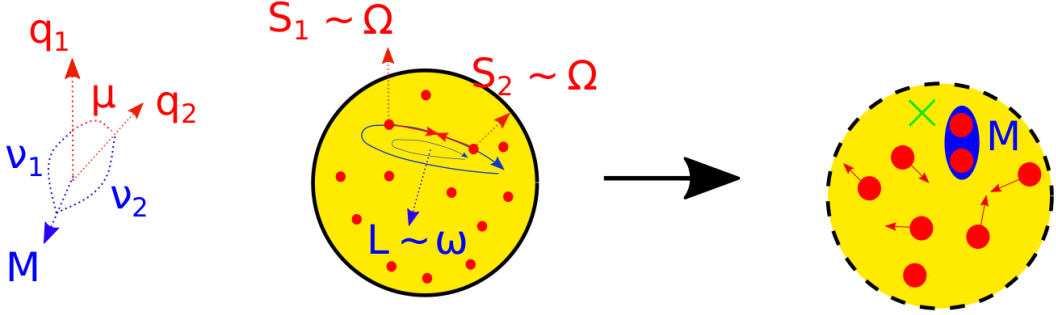}
          \caption{ Coalescence of a vector meson from two quarks, including the effect of an out-of-equilibrium spin density $\Omega$ with vorticity $\omega$.  Quark spin is determined early on and vorticity (angular momentum number) at freezeout}
                \label{coalfig}
        \end{center}
\end{figure}
\color{black}
We shall implement a coalescence picture in spin space, related to that in \cite{phicoal} but allowing vorticity and quark spin to combine in different directions to reflect the fact that spin and angular momentum are generally not in equilibrium, and are related by a diffusion equation with a characteristic timescale.

Putting in coalescence, this means $\Omega$ represents the angular momentum at the early stage, when quark angular momentum is fixed, while $L$ represents the angular momentum at hadronization, carried by the vorticity $\omega$.  Without loss of generality, therefore, $L$ can be assumed to point in the $x-z$ direction while $\vec{\hat{\Omega}}$ is a vector in the direction $z$.  (See \figref{coalfig})
\begin{equation}
  \label{omegas}
  \vec{\omega}=\omega_x\;\hat{x}+\omega_z\;\hat{z} \eqcomma \vec{\hat{\Omega}}= \Omega \hat{z}
\end{equation}
  \color{black}The estimate we make here is on the vector resonance spin density matrix only, with any momentum space dependence traced over in the co-moving frame of the resonance.  Hence, our estimate is at best a qualitative one, since it neglects the correlation between vorticity and the local velocity w.r.t. the lab frame, that would show up as ``weights'' in the transformation between the two.  We expect this issue could become important w.r.t. longitudinal flow and transverse vorticity \cite{uslisa}, and a detailed analysis of the pseudorapidity dependence of the various coefficients of \eqref{dis} is necessary to disentangle it.  
  
  Nevertheless, a non-trivial dependence of the coefficients of Eq. \ref{dis} on the lack of equilibrium between spin and vorticity angular momentum is sufficient to seriously consider these coefficients as a promising experimental observable\color{black}.

The interplay between these two angular momenta can be implemented at the wavefunction level using a coalescence formalism \cite{coal1,coal2,coal3}.  In spin space, \color{black} in the frame of reference aligned with the final vorticity $\Omega$, we can parametrize the coalesced vector meson density matrix from the spinor quark density matrices as well as coefficients that factorize the usual way (Clebsch-Gordan, 6j symbols), isolating all model dependence in a single coefficient.  The expression for the vector meson density matrix is
\begin{equation}
\label{coal}
\begin{array}{c}
\left( \hat{\rho}^M \right)_{mn} =   \sum_{ijkl} \left( P_{12}^L \right)_{ijklmn} U_S(\phi_r,\theta_r)\left( U_\omega({\mu_1,\nu_1}) \rho^{1}(\Omega) U^{-1}_\omega(\mu_{1},\nu_1) \right)_{ij}\\
\times\left( U_\omega(\mu_{2},\nu_2) \rho^{2} (\Omega) U^{-1}_\omega(\mu_2,\nu_2)\right)_{kl}U^{-1}_S(\phi_r,\theta_r)
\end{array}
\end{equation}
\color{black}
Where $M$ refers to the meson, $S_{1,2}$ are the spins of the constituent quarks and $L$ is the spatial angular momentum given by the vorticity.   \color{black}The matrices $U_\omega$ are SU(2) rotation matrices that rotate the spins in the direction of the vorticity, \textcolor{black}{depending on the relative angles $\nu_{1,2}$ of each quark's spin w.r.t.} the vorticity frame.   Physically the matrices $U$  reflect that $\Omega$ (connected to the quark spin $S_{1,2}$ quantum numbers) and $\omega$ (connected to the spacetime angular momentum quantum number $L$) are both angular momenta generated by the vortical evolution of the fluid, but they are not necessarily the same, nor they point in the same direction.

Finally, the qutrit rotation matrix $U_S$ transforms the system from the reference frame where $\omega$ is in the plane $x-z$ to the lab frame where $\theta$ and $\phi$ are defined, examined in section \ref{spinexp} (the weights discussed after \eqref{omegas} would be attached to $U_s$).

$\hat{\rho}_M$ is a vector meson $3\times 3$ density matrix, to be compared with Eq. \ref{rhogen} and $\hat{\rho}^{1,2}$ are spinor $2\times 2$ density matrices with distributions analogous to \cite{zanna,tinti},specifically for a temperature $T$ and vorticity angular momentum $\hat{\Omega}$  at the {\em early} stage when quarks are polarized (not the final vorticity at freezeout).   Since quarks are close to local equilibrium when polarized
\textcolor{black!75!black}{\begin{equation}
  \label{rhoquark}
\hat{\rho}^{1,2}_{kl} = \frac{1}{Z}\exp \left[  \frac{\vec{\hat{\Omega}}\cdot\vec{\hat{\sigma}}}{2T} \right]
  \end{equation}
  }
with the usual normalization $\hat{\rho}_{kk}=1$ and $\sigma$ the usual Pauli spin matrices.

\color{black}  $P^L_{12}$ are the coefficients of the coalescence Wigner function projected into spin space.  They, and even their normalization, are only partially calculable since they incorporate all configuration space and momentum space dynamics.   However, the fact that this dynamics is rotationally invariant allows us to use the Wigner-Eckart theorem to parametrize these unknowns into only a few normalization coefficients.  

Assuming rotational invariance in the meson rest frame, coalescence will project an eigenstate which in spin space is equivalent to the addition of 3 angular momenta $s_1,s_2,L$ for which one can use Wigner 6j symbols.   We shall assume, a reasonable guess in line with coalescence, that first the hadron wavefunction is formed ($s_1$ and $s_2$ are added) and then it aligns with $L$. \color{black} Hence \cite{pratt}, by the Wigner-Eckart theorem the array $P_{12}^L$ in \eqref{coal} is identified with
\begin{equation}
  \label{cleb}
P_{12}^L \equiv \left( C_{S_{1} S_{2}}^L \right)_S=\sum_{m_1,m_2,m_L,m_{12}} P_L(\omega) C^{S_1+S_2,L}_{m_{12},m_L,S,M_S} C^{S_1,S_2}_{m_1,m_2,S_1+S_2,m_{12}}
\end{equation}
where $C^{...}_{...}$ are the Clebsh-Gordon coefficients and index-counting matches Eq. \ref{coal} and $P_L(\omega)$ is the probability of a thermalized fluid with vorticity $W$ to transfer momentum quantum number $L$ to the coalescence Wigner function.  This we do not know anything about, beyond the reasonable expectation that larger $\omega$ will favor higher $L$ and the fact that angular momentum conservation will only allow $L=0,1,2$ to be transferred to a meson of spin 1.

After some algebra, described in \appref{cg} and based on the same rotation operations described in the last section (\eqref{dmat} and \eqref{dexp}), we arrive at the meson density matrix, which is given by:
\textcolor{black!75!black}{
	\begin{equation}
	\left( \hat{\rho}^M \right)_{m',m'''}= \int_0^{2\pi} P_W(\nu) d\nu \sum_{m'' , m, \tilde{m}'', \tilde{m}}P^{2}_L(\omega) \tilde{C}^{L}_{m^{''},m} e^{i(m'''-m')\phi_r}d^{j}_{m' \tilde{m}''}(\theta_r)d^{j}_{\tilde{m}'' m^{''} }(\nu)\times
	\label{reDensity}
	\end{equation}
	\[\times\ket{j,m''}\rho^{1}(\Omega)\rho^{2} (\Omega)\bra{j,m}\left[d^{j}_{m \tilde{m}}(\nu)\right]^{-1}\left[d^{j}_{\tilde{m} m'''}(\theta_r)\right]^{-1}
	\]}
with each coefficient given explicitly in \appref{cg} as well as \eqref{dmat} and \eqref{dexp} and $P_W(\nu)$ is the angular Wigner function of two quarks aligned w.r.t. angle $\nu$ to coalesce.
\textcolor{black}{Note that the $L=0$ with $\theta_r=0$, $\phi_r=0$ and $P_W(\nu)\rightarrow \delta(\nu)$ case reduces to the one examined in \cite{xia}, as we explicitly show in appendix \ref{appxia}.}

We can now calculate the distribution of coefficients of Eq. \ref{tablecoeff}
in the case that spin and angular momentum are {\em not} necessarily aligned, as expected in \cite{uscausality}, via  Eq. \ref{rhogen} and Eq. \ref{cleb} and \ref{coal}.  Without loss of generality, we define the plane defined as $\nu$ as $\phi_r=0$.
The full expressions are rather long, and we have put them in \appref{seccoal}, assuming a Gaussian Wigner function $P_W(\nu)$.
The main message is the sensitivity of the observables of the array in  \eqref{tablecoeff} to {\em both} quark polarization and vorticity within a coalescence scenario.

\begin{figure}[h]
		\begin{center}
			\epsfig{width=0.49\textwidth,figure=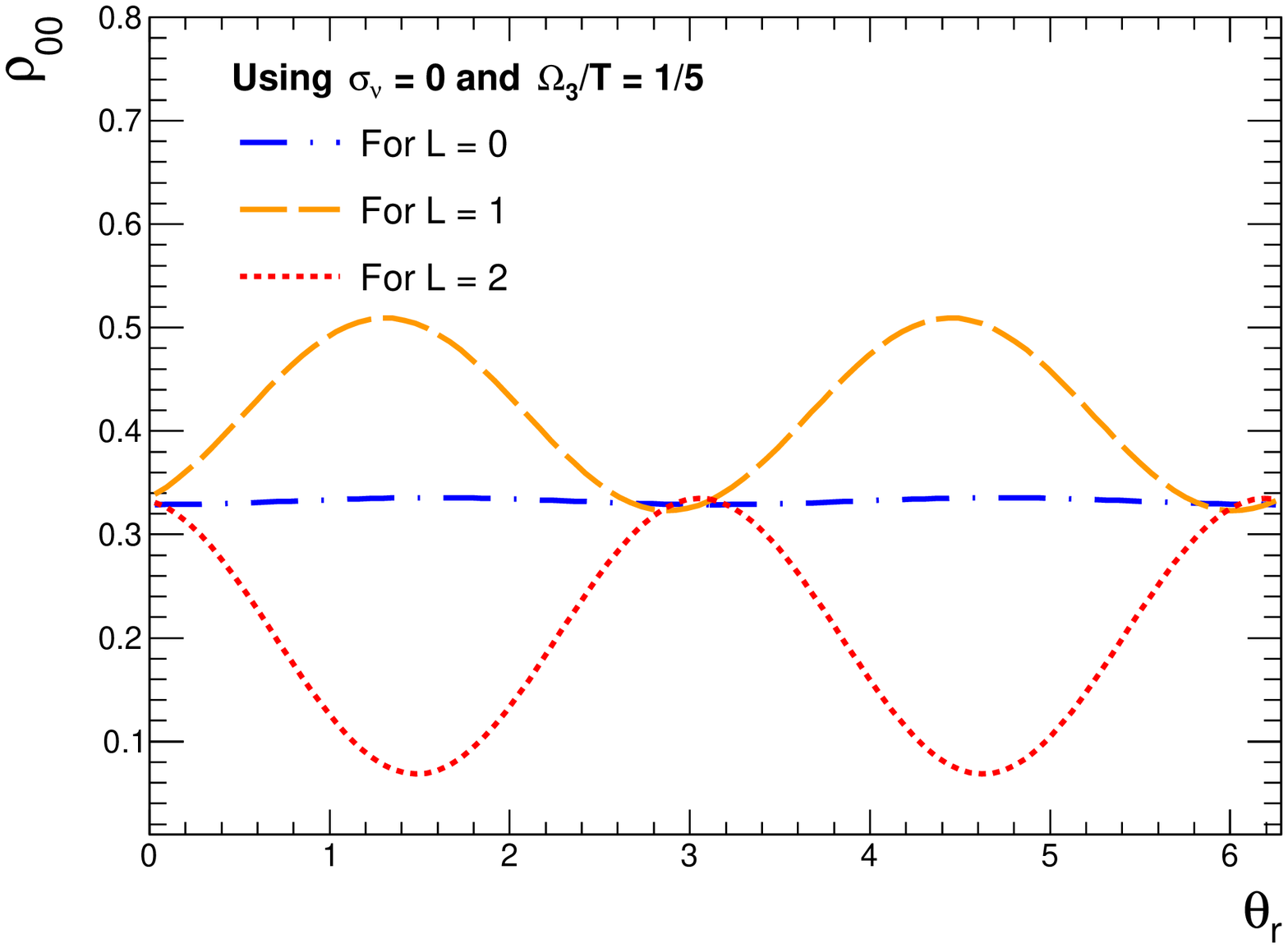}
			\epsfig{width=0.49\textwidth,figure=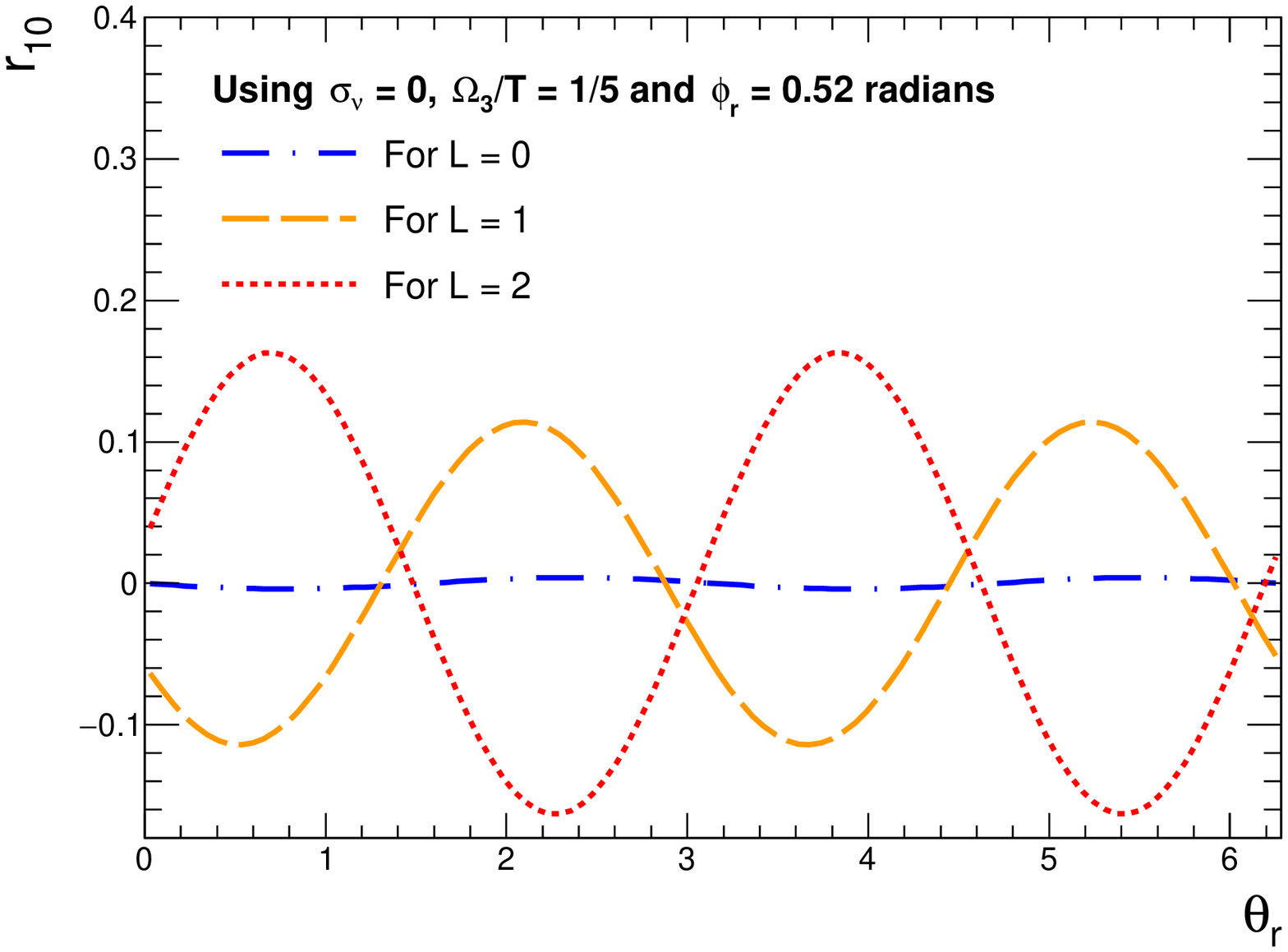}
			\epsfig{width=0.49\textwidth,figure=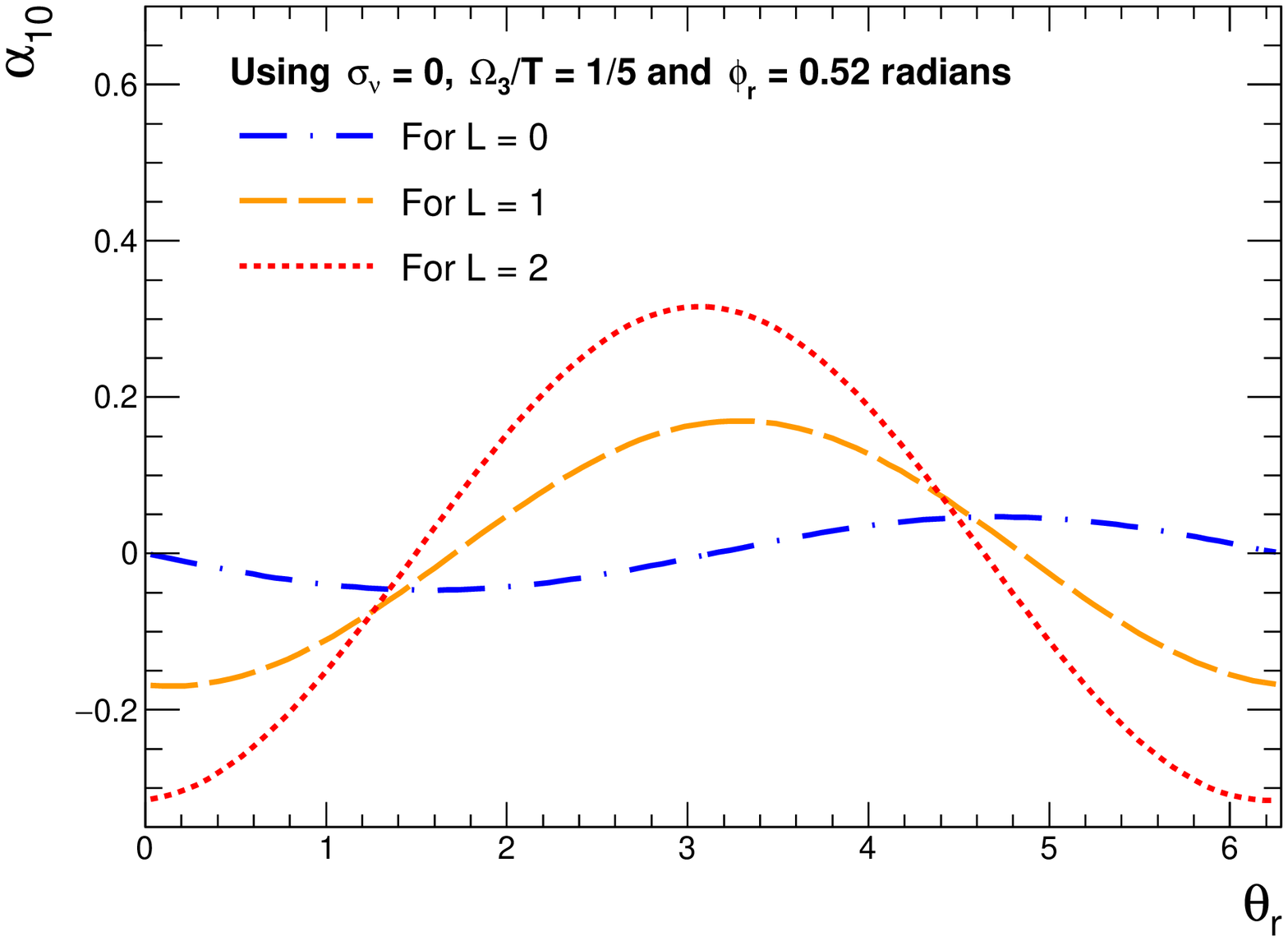}
			\epsfig{width=0.49\textwidth,figure=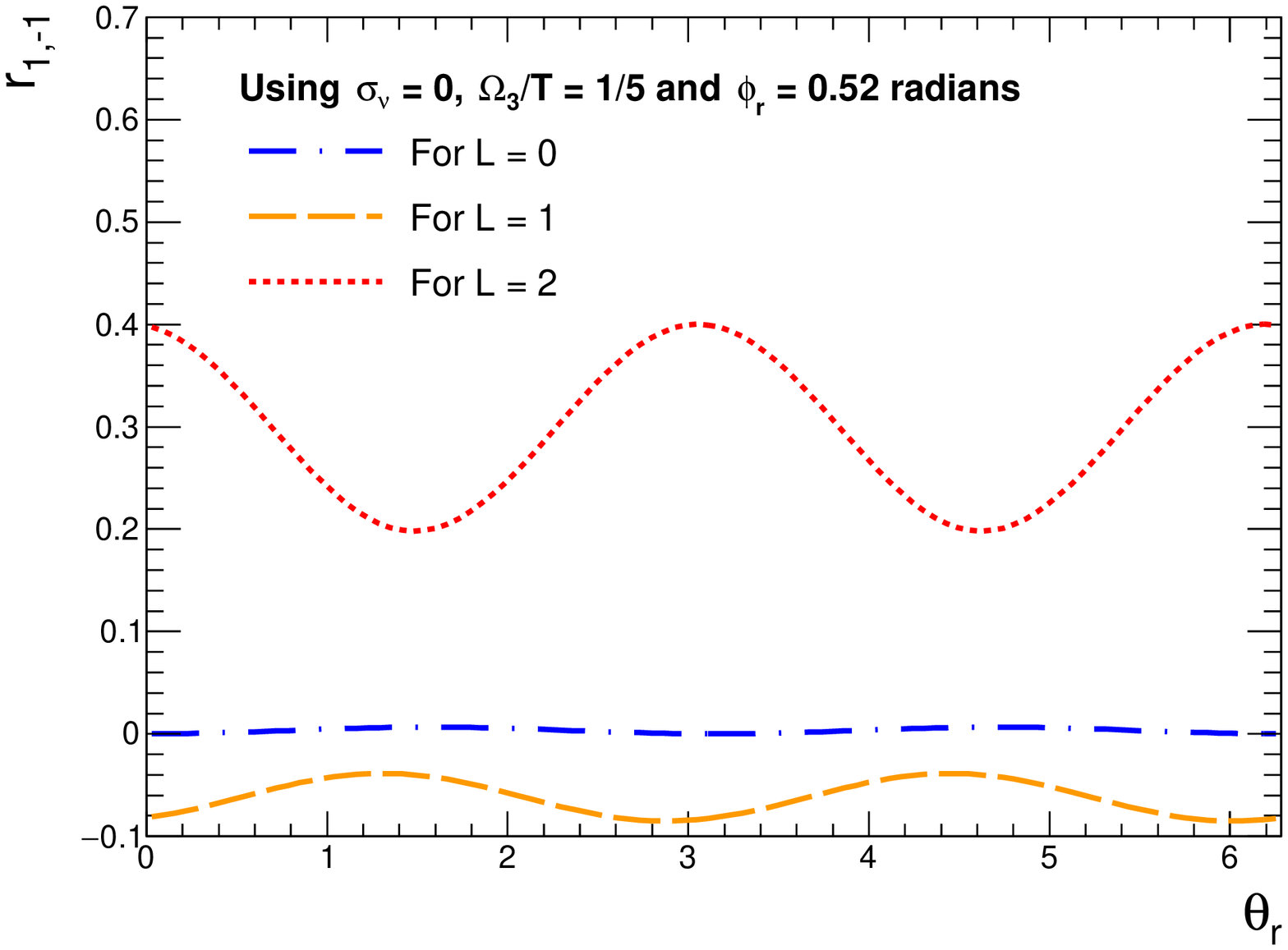}
			\epsfig{width=0.49\textwidth,figure=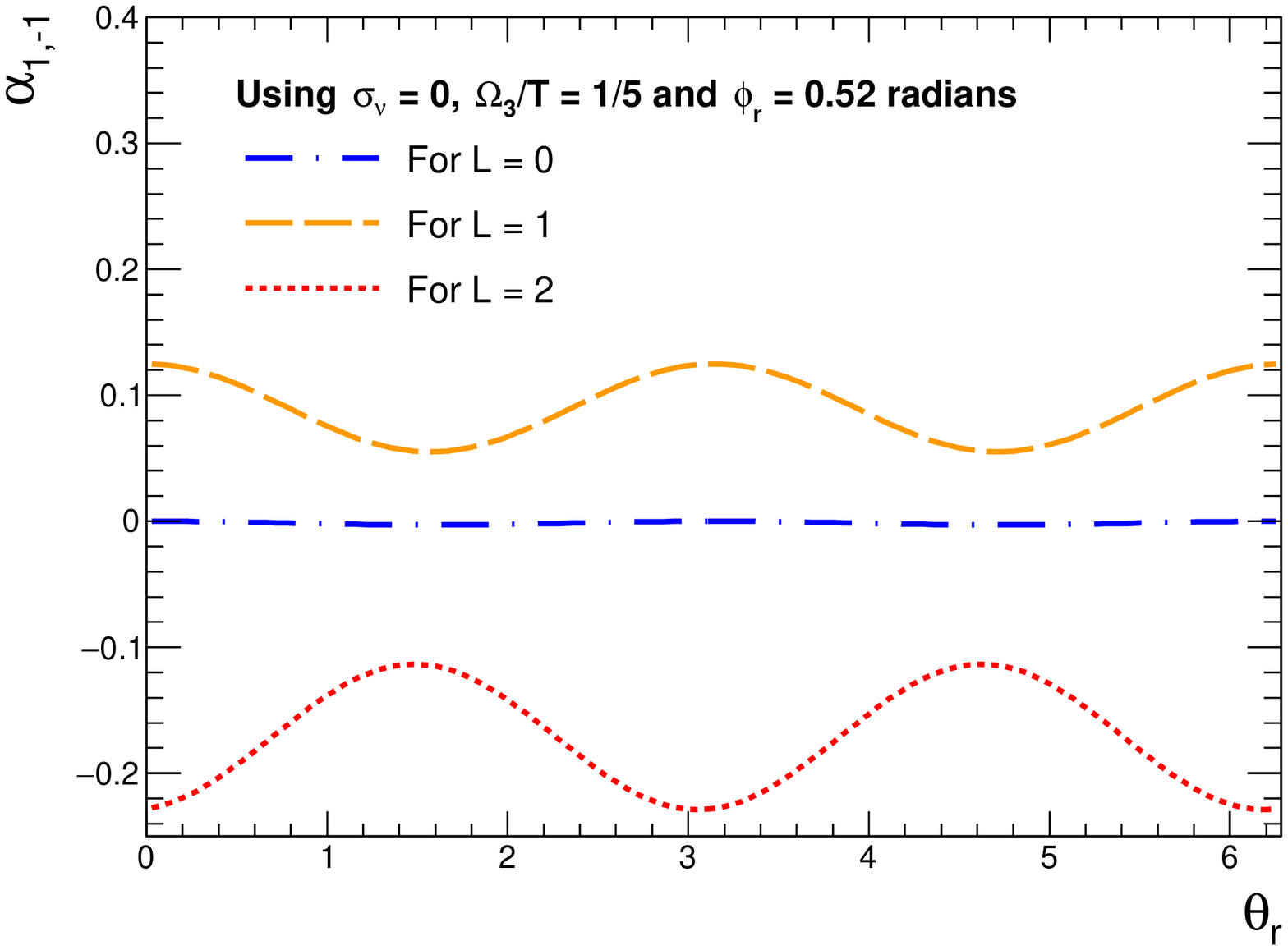}
\end{center}
                        \caption{     \label{fignoneq} Coefficients of \eqref{dis}calculated in the coalescence model for a representative choice of parameters.  The $L=0,1,2$ represents the angular momentum number the meson obtains from the vorticity, something that we can not calculate within the current formalism}
	\end{figure}

To illustrate this, we choose a set of parameters that fit $\rho_{00}$ close to $1/3$ in the ``trivial'' $L=0$ limit, $\sigma_\nu=0$, $\Omega_3/T = 1/5$ and $\phi_r = 0.5$, with negligible $\sigma_\nu$.   We obtain \figref{fignoneq}. We can see from figures that both $L=1$ and $L=2$ contributions for coefficients $\rho_{00}$ and $r_{10}$ oscillate, with opposite signs.
$\alpha_{1,-1}$'s width changes, and the baseline amplitudes of $\alpha_{10}$ and $r_{1,-1}$ are very different for the different $L$ contributions.   All these coefficients are therefore potentially sensitive probes of spin-angular momentum non-equilibrium.
\textcolor{black!75!black}{The plots are not meant to be quantitative, this will depend sensitively on the details of the hydrodynamic evolution, vorticity formation, coalescence Wigner function etc.
Rather, they are indicative of which component is most sensitive to non-equilibrium between spin and vorticity.}

These equations would need to be integrated over probabilities to get $\theta_r,\phi_r$ as a function of $\Omega$ and $w$ and the momentum region, which need to be read off a hydrodynamic model.  Initial vorticity going into quark spin, $\Omega_{1,2,3}$ and $T$, as well as final vorticity $\omega$, would also need to be obtained from such a model.

Summarizing this section, we are now in a position to get some information on the coefficients of Eq. \ref{dis} based on the dynamics of the system.    The requirement of rotational invariance, implemented by Wigner function techniques, gives us quite a lot of, although not quite the full information linking spin 1 spin alignment to the evolution of angular momentum and spin currents.
What is the angular momentum corresponding to this vorticity, parametrized by $P_L(\omega)$ of course is not known, but all coefficients are proportional to it in the same way.  Since one expects higher vorticity to transmit more angular momentum to the hadron it is expected to be monotonic in $\omega,L, \sim w^{\alpha_1}L^\alpha_2,\alpha_{1,2} \geq 0$.
\color{black}

The figures in this section indicate which Fourier coefficients of Eq. \ref{dis} are more sensitive to non-equilibrium between spin and vorticity.  By eye, $r_{1,0},r_{1,-1}$ and $\alpha_{1,-1}$ are approximately equally constraining.   But what is sure is that a simultaneous fit of the coefficients of \eqref{tablecoeff} and their dependence on rapidity and reaction plane angle have the potential of thoroughly constrain the degree of non-equilibrium between vorticity and polarization.
\section{Discussion \label{discuss}}
Eq. \ref{dis} with Eq. \ref{rhogen}, make it clear that, given a good distribution of $\theta,\phi$ and given an non-chaotic variation of $\theta_r,\phi_r$ between events, one should be able to recover a distribution of $n_{3,8}$.  It is a matter of Fourier analysis of Eq. \ref{dis} and algebra.

In \secref{spinexp} we showed that this is indeed the expectation, and computed a complicated but closed form function relating experimental observables to the spin density matrix, \eqref{n3eqdef} and \eqref{n8eqdef}
and, crucially, the rotation angles $\theta_r,\phi_r$ (given respectively by \eqref{phirdef} and \eqref{thetardef}) between between the experimentally accessible frame and one optimized w.r.t. the qutrit parametrization of \eqref{rho8} 
These, if centered around a particular region, can give information about the spin-hydrodynamic evolution, and crucially the degree of coherence of the spin wavefunction of the meson.

What can we learn from all this?   Here we have to get a bit more
qualitative, with two limiting cases.
\begin{description}
\item[(i) Cooper-Frye freezeout]
If spin and vorticity are in equilibrium, one expects that Cooper-Frye
is a good estimate.   This means that the spin of this meson is not a
coherent state, but rather a superposition of Eigenstates of the type of \eqref{maxmixed}.
As Eq. \ref{phirdef} makes clear, in this case $\phi_r$ should have a uniform event-by-event distribution.
\item[(ii) Wigner function coalescence]
If spin and vorticity are not in equilibrium, of course tings become a
bit more model-dependent.     But a coalescence formalism (see
conclusion of \cite{uscausality} ) means the matrix is a pure state for some choice of $\theta_r,\phi_r$.   It is a good guess that One of these angles is related to vorticity and the other to spin. \textcolor{black!75!black}{In this case we could make a direct link to the non-equilibrium between vorticity and polarization from the analysis of the coefficients in \eqref{tablecoeff} as illustrated in \figref{fignoneq}.}
\end{description}
As an immediate test, it is worth to check if $\phi_r$ in Eq. \ref{phirdef} has a peaked distribution event-by-event, between the production plane and a beam plane.   As argued at the end of \secref{spinexp}, this would be evidence that freeze-out is happening dynamically, rather than by Cooper-Frye.  Similar points were made in a recent work \cite{xia}.

A further more involved analysis would be a purity test on $\rho$, i.e. to verify experimentally if
 $  \rho=\rho^2$.
For Eq. \ref{rho8} this is tantamount to verifying that for some choice of $\theta_r,\phi_r$ we have $\left(n_3,n_8\right)=(0,-1),\;\frac{1}{2}(\pm\sqrt{3},1)$\cite{qutrit,geom}.
  One also needs to check the relation with $\Lambda$ polarization, in other words that the $\theta_r,\phi_r$ obtained from the analysis in section \ref{spinexp} are also close to the mean polarization axis of the $\Lambda$ baryon (see  eq. 2-3 of \cite{oliva}).   \eqref{thetardef} and \eqref{phirdef} make it clear that this will happen if both $\theta_r$ and $\phi_r$ are well-defined and centered around zero.  This indicates spin and angular momentum are in relative equilibrium, but a non-Cooper-Frye freezeout (For Cooper-Frye the distribution of $\phi_r$ would need to be relatively homogeneous)

  A maximally impure state \eqref{maxmixed} is a signature of a Cooper-Frye freeze-out, where indeed spin and orbit are fully in equilibrium.   Conversely, purity will indicate that freeze-out dynamics is more complicated than that, as indeed causality  \cite{uscausality} and the fact that hydrodynamics with spin cannot be derived purely from conservation laws \cite{tinti} seem to indicate.

The simplest hadronization model besides Cooper-Frye is coalescence. In section \ref{spincoal} we used this model, together with considerations coming from rotational symmetry, to understand how the coefficients derived in this section relate to the spin and vorticity of the quark-gluon plasma. A generic expectation is that $\theta_r$ and $\phi_r$ are indeed  determined by the interplay of primordial spin current and developed vorticity current, not necessarily in equilibrium with each other.

\color{black} The techniques developed in the second part of this work allow us to relate the decay matrix elements of Eq. \ref{dis} to theoretical calculations where spin and angular momentum are not in equilibrium, such as \cite{uscausality}.  Currently no numerical implementation exists, although analytical calculations look quite promising, as qualitatively they agree with global $\Lambda$ polarization \cite{analspin2}.   Weather lack of equilibrium between spin and vorticity can indeed account for the apparent inconsistency between spin alignment and Fermion polarization will be left to a forthcoming work, \color{black} since it would require understanding coalescence and decay dynamics in the co-moving frame as well as a quantitative understanding of the flow profile throughout the fireball (to transform all angles to the lab frame accurately). \color{black}

Another obvious direction is to extend the current formalism to Baryons. Doing so will determine if the current formalism can explain the experimental puzzle that at LHC energies $\Lambda$ polarization is consistent with zero, while spin alignment is not  \cite{alice}. A look at Equation \ref{coal} shows such an explanation is indeed possible:
The equivalent of \eqref{coal} for baryons would be a yet more
complicated coalescence term on the right hand side (3 fermions and a vorticity) and a yet more restrictive wavefunction on the left hand side (a qubit instead of a qutrit). Hence, there are more ways for vorticity and quark spin effects to cancel in a baryon rather than a spin 1 meson. 

In conclusion, vector spin alignment is an interesting observable which has not to date been fully exploited.   A full harmonic analysis of the vector meson's decay coefficients \eqref{tablecoeff} would have the potential to elucidate the mechanism determining hadron spin at freeze-out and also to provide experimental evidence of lack of equilibrium between spin and angular momentum.   We eagerly await experimental data in this direction.

GT thanks CNPQ bolsa de produtividade 306152/2020-7, bolsa FAPESP 2021/01700-2 and participation in tematic FAPESP, 2017/05685-2. 
K.J.G. is supported by CAPES doctoral fellowship 88887.464061/2019-00

\appendix
\section{Relation between Gell-Mann matrix elements and coefficients of \eqref{dis}\label{seclambda}}
The elements of the density matrix are parametrized according to \eqref{sigma3def} as well as the angle $\theta_r$ in terms of Eq. \ref{dis} are
\begin{equation}
  \label{n3eqdef}
	n_3\left(\rho_{00},r_{10},\phi_r\right) = \frac{1}{8 r_{10}^2}\sqrt{\frac{3}{2}} \alpha _{10} \csc
	\left(\phi _r\right) \;\lambda_2\left(\rho_{00},r_{10},\phi_r\right)
	\left((1-3 \rho_{00})^2+2 \;\lambda_1\left(\rho_{00},r_{10},\phi_r\right)\right.+
\end{equation}
\[\left.+(1-3
\rho_{00})^2 \cos \left(2 \phi _r\right)+12
r_{10}^2\right)+\frac{\sqrt{3}}{24 \rho_{00}-8}\;\left((1-3
	\rho_{00})^2-3 \sec ^2\left(\phi _r\right)
	\lambda_1\left(\rho_{00},r_{10},\phi_r\right)\right)
\]
\begin{equation}
  \label{n8eqdef}
n_8\left(\rho_{00},r_{10},\phi_r\right)= \frac{3 \alpha _{10} \csc \left(\phi
	_r\right)}{16 r_{10}^2}\left((1-3 \rho_{00})^2+2\;\lambda_1\left(\rho_{00},r_{10},\phi_r\right)+(1-3 \rho_{00})^2 \cos \left(2 \phi
	_r\right)+\right.
\end{equation}
	\[\left.+12 r_{10}^2\right)\;\lambda_3\left(\rho_{00},r_{10},\phi_r\right)+\frac{1}{8} \left(-3
\rho_{00}+\frac{3 \sec ^2\left(\phi _r\right)
	\;\lambda_1\left(\rho_{00},r_{10},\phi_r\right)}{3 \rho_{00}-1}+1\right)
\]
\begin{equation}
\label{thetardef}
  \theta_r\left(\rho_{00},r_{10},\phi_r\right)= \tan
^{-1}\left(\frac{\lambda_4\left(\rho_{00},r_{10},\phi_r\right)}{\lambda_5\left(\rho_{00},r_{10},\phi_r\right)}\right)
\end{equation}
where the functions $\lambda_i$ are
\begin{equation}
\lambda_1\left(\rho_{00},r_{10},\phi_r\right) = \sqrt{(1-3 \rho_{00})^4 \cos ^4\left(\phi _r\right)+4 (1-3 \rho_{00})^2 r_{10}^2 \cos ^2\left(\phi _r\right)}
\end{equation}
\begin{equation}
\lambda_2\left(\rho_{00},r_{10},\phi_r\right) = \sqrt{\frac{r_{10}^2}{(1-3
		\rho_{00})^2 \cos ^2\left(\phi
		_r\right)+\lambda_1\left(\rho_{00},r_{10},\phi_r\right)+6 r_{10}^2}}
\end{equation}
\begin{equation}
\lambda_3 \left(\rho_{00},r_{10},\phi_r\right)= \sqrt{\frac{(1-3
		\rho_{00})^2-\sec ^2\left(\phi _r\right)
		\left(\lambda_1\left(\rho_{00},r_{10},\phi_r\right)-6 r_{10}^2\right)}{4 (1-3
		\rho_{00})^2+18 r_{10}^2 \sec ^2\left(\phi
		_r\right)}}
\end{equation}
\begin{equation}
\lambda_4\left(\rho_{00},r_{10},\phi_r\right) = -\frac{\sec \left(\phi _r\right)}{(3 \rho_{00}-1)
	r_{10}} \left((1-3
	\rho_{00})^2+2 \lambda_1\left(\rho_{00},r_{10},\phi_r\right)+\right.
\end{equation}	
	\[\left.+(1-3 \rho_{00})^2 \cos \left(2 \phi _r\right)\right)
	\lambda_3\left(\rho_{00},r_{10},\phi_r\right)\]
\begin{equation}
\lambda_5\left(\rho_{00},r_{10},\phi_r\right) = -4 \sqrt{\frac{(1-3 \rho_{00})^2 \cos
		^2\left(\phi _r\right)-\lambda_1\left(\rho_{00},r_{10},\phi_r\right)+6 r_{10}^2}{2
		(1-3 \rho_{00})^2 \cos ^2\left(\phi _r\right)+9
		r_{10}^2}}
\end{equation}
\section{Details of the density matrix calculations \label{cg}}
\begin{figure}[h]
	\begin{center}
		\includegraphics[width=0.49\linewidth]{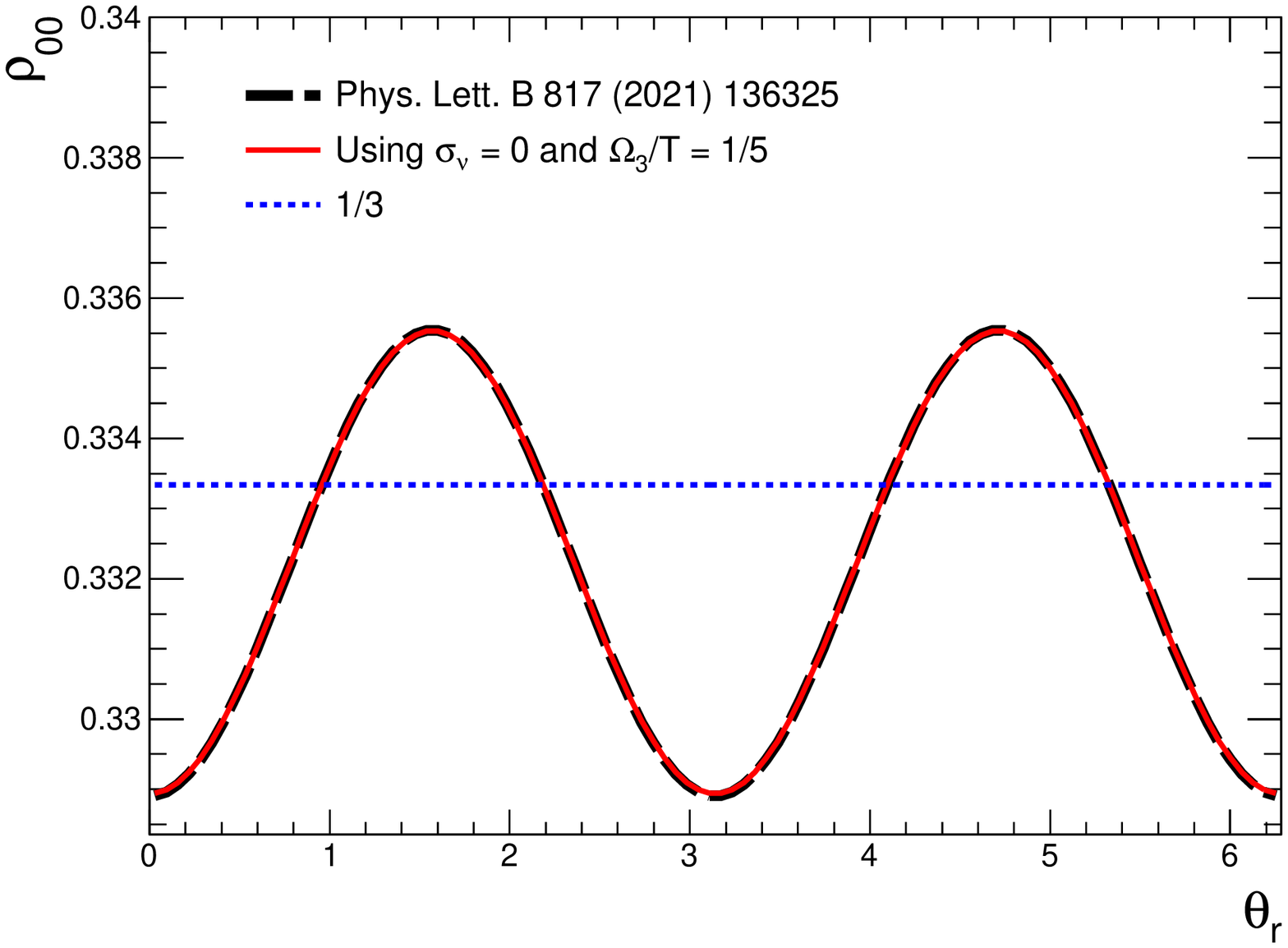}
		\includegraphics[width=0.49\linewidth]{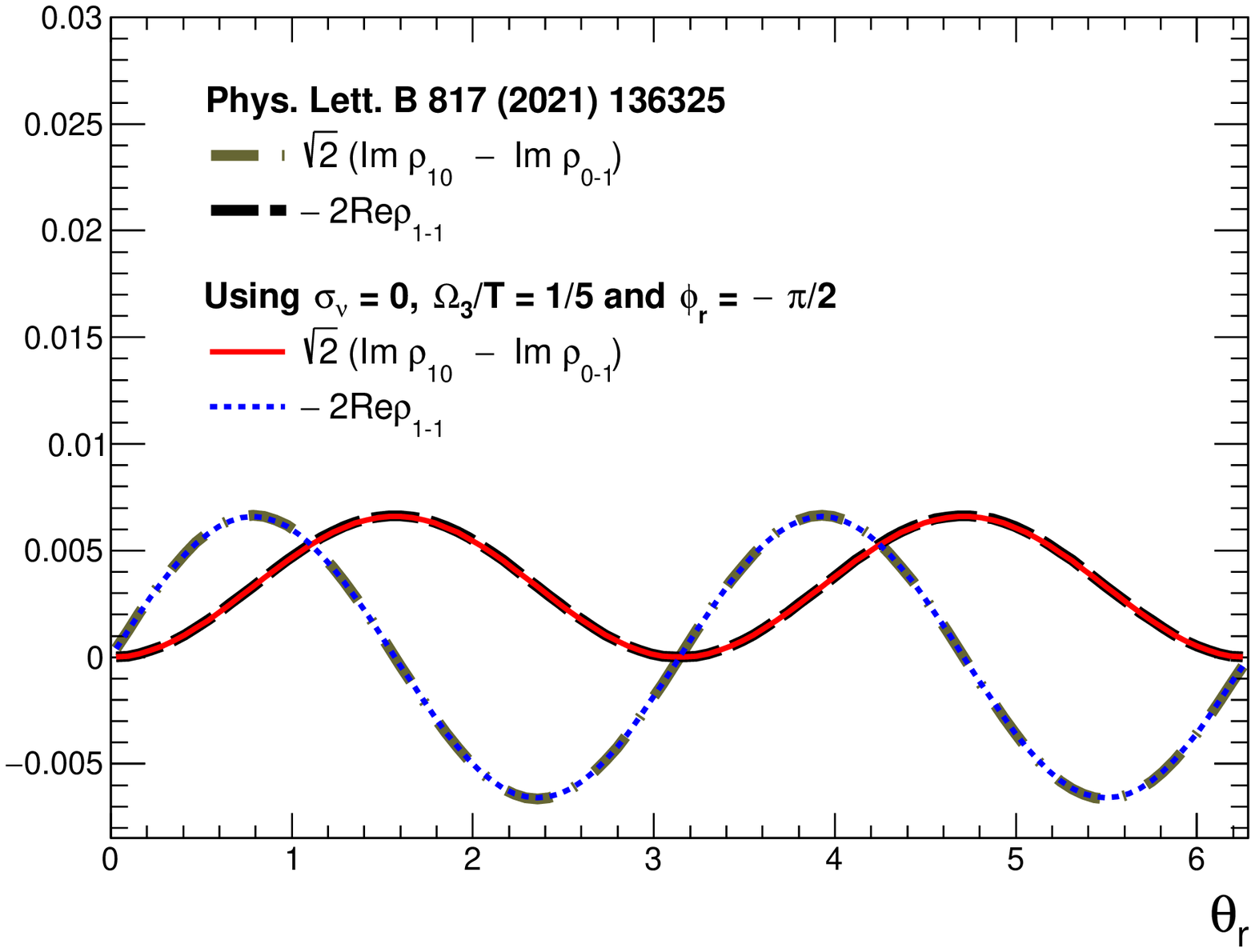}
		\caption{\label{xiaappplot} Comparison of our model with the one in \cite{xia} in the appropriate limit}
	\end{center}
\end{figure}

From the density matrix for mesons \eqref{coal}, the rotation matrix composition explicitly form is
\begin{equation}
  \label{matrices}
  U_S(\phi_r,\theta_r)U_\omega({\mu_1,\nu_1}) \rho^{1}(\Omega) \times U^{-1}_\omega(\mu_{1},\nu_1) U_\omega(\mu_{2},\nu_2) \rho^{2} (\Omega) U^{-1}_\omega(\mu_2,\nu_2)U^{-1}_S(\phi_r,\theta_r)
\end{equation}
  .
Rotational symmetry allows us to approximate $\mu_{1} \simeq \mu_{2} \simeq \mu$ and $\nu_{1}\simeq \nu_{2} \simeq \nu$ and put the residual uncertainty in the smearing width (it appears as $\sigma_\nu$ in \eqref{nugauss}).   Thus \eqref{matrices} reduce to
\[
	= U_S(\phi_r,\theta_r)U_\omega({\mu,\nu}) \rho^{1}(\Omega) U^{-1}_\omega(\mu,\nu) U_\omega(\mu,\nu) \rho^{2} (\Omega) U^{-1}_\omega(\mu,\nu) U^{-1}_S(\phi_r,\theta_r)
\]
\[
	=U_S(\phi_r,\theta_r)U_\omega({\mu,\nu}) \rho^{1}(\Omega)\rho^{2} (\Omega) U^{-1}_\omega(\mu,\nu)U^{-1}_S(\phi_r,\theta_r)
\]
we can now apply \eqref{dmat} and \eqref{dexp} to reduce these equations to matrix form, and use \eqref{rhogen2} to expand
\[
=\sum_{m'' , m, \tilde{m}'', \tilde{m}} \ket{j,m'}D_S(\phi_r,\theta_r)\bra{j,\tilde{m}''}\ket{j,\tilde{m}''}D_\omega\left(\mu,\nu\right)\bra{j,m''}\ket{j,m''}\rho^{1}(\Omega)\rho^{2} (\Omega)\bra{j,m}\times\]
\[\times\ket{j,m}D_\omega\left(\mu,\nu\right)^{-1}\bra{j,\tilde{m}}\ket{j,\tilde{m}}D^{-1}_S(\phi_r,\theta_r)\bra{j,m'''}
\]
\[
=\sum_{m'' , m, \tilde{m}'', \tilde{m}} e^{i(m'''-m')\phi_r}e^{i(\tilde{m}-\tilde{m}'')\mu}d^{j}_{m' \tilde{m}''}(\theta_r)d^{j}_{\tilde{m}'' m^{''} }(\nu)\times\]
\[\times\ket{j,m''}\rho^{1}(\Omega)\rho^{2} (\Omega)\bra{j,m}\left[d^{j}_{m \tilde{m}}(\nu)\right]^{-1}\left[d^{j}_{\tilde{m} m'''}(\theta_r)\right]^{-1}
\]
Making 
\begin{equation}
	\label{t}
		T_{m',m'''}(\theta_r,\nu,\Omega)=d^{j}_{m' \tilde{m}''}(\theta_r)d^{j}_{\tilde{m}'' m^{''} }(\nu)\ket{j,m''}\rho^{1}(\Omega)\rho^{2} (\Omega)\bra{j,m}\left[d^{j}_{m \tilde{m}}(\nu)\right]^{-1}\left[d^{j}_{\tilde{m} m'''}(\theta_r)\right]^{-1}
\end{equation}
\[
	= \sum_{m'' , m, \tilde{m}'', \tilde{m}}e^{i(m'''-m')\phi_r}e^{i(\tilde{m}-\tilde{m}'')\mu}T_{m',m'''}(\theta_r,\nu,\Omega)
\]
\textcolor{black!75!black}{
	So,  from the Clebsch Gordan coefficients \ref{cleb} and making $M'_S=m^{''}$ and $M_S=m$, we can write the following equation:
	\small\begin{equation}
	\tilde{C}^{L}_{m^{''},m}= \sum_{m'_1,m'_2,m'_L,m'_{12}}\;\sum_{m_1,m_2,m_L,m_{12}}\;C^{S_1+S_2,L}_{m'_{12},m'_L,S,M'_S} C^{S_1,S_2}_{m'_1,m'_2,S_1+S_2,m'_{12}}\; C^{S_1+S_2,L}_{m_{12},m_L,S,M_S} C^{S_1,S_2}_{m_1,m_2,S_1+S_2,m_{12}}
	\end{equation}
}
So 
\textcolor{black!75!black}{\begin{equation}
	\left( \hat{\rho}^M \right)_{m',m'''} = \sum_{m'' , m, \tilde{m}'', \tilde{m}}\;P^{2}_L(\omega)\; \tilde{C}^{L}_{m^{''},m} e^{i(m'''-m')\phi_r}e^{i(\tilde{m}-\tilde{m}'')\mu}T_{m',m'''}(\theta_r,\nu,\Omega)
\end{equation}}
putting all components of $\rho_M$ together \eqref{reDensity} follows
\color{black}{	\section{Comparison of our calculations with pre-existing models( \cite{xia}) \label{appxia}}
	From equation \ref{rho00} for the case where $L = 0$ and $\sigma_\nu = 0$, we have, after some algebra
	\small\begin{equation}
	\langle\rho^{L=0}_{00}(\theta_r,\Omega_3,T)\rangle
	= \frac{\tilde{P}^{2}_{L}\left(\omega\right)}{2}
	\left(1+\cosh\left(\frac{\Omega _3}{T}\right)-\left(\cosh\left(\frac{\Omega _3}{T}\right)-1\right)\cos\left(2\theta_r\right)
	\right)
	\end{equation}
	using $\tilde{P}^{2}_{L}\left(\omega\right)=\frac{1}{1+2\cosh\left(\frac{\Omega _3}{T}\right)}$, this simplifies to 
	\[
\langle\rho^{L=0}_{00}(\theta_r,\Omega_3,T)\rangle	=\frac{1+\cosh\left(\frac{\Omega _3}{T}\right)}{2\left(1+2\cosh\left(\frac{\Omega _3}{T}\right)\right)}\left[1-\left(\frac{\cosh\left(\frac{\Omega _3}{T}\right)-1}{\cosh\left(\frac{\Omega _3}{T}\right)+1}\right)\cos\left(2\theta_r\right)\right]
	\]
	Then one can make a  comparison with the $\rho_{0,0}$ coefficient as calculated in \cite{xia} (equation 28-29)
	\begin{equation}
	\rho_{0,0}\left(\Delta\Psi\right) = \frac{1-F^2_{\perp}\cos\left(2\Delta\Psi\right)}{3+F^2_{\perp}}
	\end{equation}
	Since the variables $\Omega_3$, $T$ are linked with \cite{xia}'s $F_{\perp}$ via
	\begin{equation}
      F^2_{\perp}    =\tanh^2\left(\frac{\Omega _3}{2T}\right)
	\end{equation}
	and it is possible to identify $\theta_r$with the angle used as an azimuthal correlation in \cite{xia}, 
	$\theta_r\rightarrow\Delta\Psi$
	For the case $r_{1,-1}$
	\begin{equation}
	-2\langle r^{L=0}_{1,-1}(\theta_r,\Omega_3,T)\rangle = \left(\frac{-2\cos\left(2\phi_r\right)\sinh^2\left(\frac{\Omega_3}{2T}\right)}{1+2\cosh\left(\frac{\Omega_3}{T}\right)}\right)\sin^2\left(\theta_r\right)
	\label{d4}
	\end{equation}
	\begin{equation}
	\sqrt{2}\left(\text{Im}\rho_{1,0}-\text{Im}\rho_{0,-1}\right)=\left(\frac{-2\sin\left(\phi_r\right)\sinh^2\left(\frac{\Omega_3}{T}\right)}{1+2\cosh\left(\frac{\Omega_3}{T}\right)}\right)\sin\left(2\theta_r\right)
	\label{d5}
	\end{equation}
putting everything together
	\begin{equation}
	-2 r_{1,-1} = \frac{2F^2_{\perp}}{3+F^2_{\perp}}\sin^2\left(\Delta\Psi\right)
	\end{equation}
	\begin{equation}
	\sqrt{2}\left(\text{Im}\rho_{1,0}-\text{Im}\rho_{0,-1}\right) = \frac{2F^2_{\perp}}{3+F^2_{\perp}}\sin\left(2\Delta\Psi\right)
	\end{equation}
	Then, from equations \ref{d4} and \ref{d5} and
  $\phi_r = -\frac{\pi}{2}$ we can reproducible the value of paper \cite{xia} as is shown in  \figref{xiaappplot}.
}

\textcolor{black}{
For the case we have $\phi_r = -\pi/2$ the coefficient $\rho_{0,0}$ is the same as shown in figure \ref{fignoneq} and other coefficients $r_{10}$ and $\alpha_{1,-1}$ are equal to zero.
}

\section{Explicit form of the distribution coefficients in a coalescence model \label{seccoal}}
Here we give form for the coefficients in the coalescence scenario.   As explained in the main text, we can not determine the angular momentum quantum number $L$ transferred from the macroscopic vortex to the vector meson.  In the main text it is paramterized by $P_L$.

The point is to write in closed form \eqref{reDensity}, and integrate over the coalescence angle $\nu$ weighted by a Wigner function $P_W(\nu)$.
$\nu$ is a parameter of the coalescence model, representing the angle of the quarks w.r.t. the meson.   Assuming the Wigner function in azimuthal space is a Gaussian of width (depending on the internal dynamics) $\sigma_\nu \ll 2\pi $
\begin{equation}
  \label{nugauss}
\textcolor{black}{P_W(\nu) = \frac{1}{\sqrt{2 \pi \sigma^2_\nu}} \exp\left[ -\frac{\nu^2}{2\sigma_\nu^2} \right]}  \eqcomma \int_0^{2\pi} P_W(\nu) f(\nu)d\nu \simeq \int_{-\infty}^{\infty} P_W(\nu) f(\nu)d\nu
  \end{equation}
  Averaging over a Gaussian Wigner function of coalesce angles, \eqref{nugauss} gives these coefficients:
\begin{description}
\item[The coefficient] $\rho_{00}$
\small\begin{equation}
\langle\rho^{L=0}_{00}(\theta_r,\Omega_3,T)\rangle=\frac{\tilde{P}^{2}_{L}\left(\omega\right)}{4}
e^{-2 \sigma_{\nu}
	^2-\frac{\Omega _3}{T}} \left(e^{2 \sigma_{\nu} ^2}
\left(e^{\frac{\Omega
		_3}{T}}+1\right){}^2-\left(e^{\frac{\Omega
		_3}{T}}-1\right){}^2 \cos \left(2
\left(\theta
_r\right)\right)\right)
\label{rho00}
\end{equation}	
\small\begin{equation}
\langle\rho^{L=1}_{00}(\theta_r,\Omega_3,T)\rangle=\frac{\tilde{P}^{2}_{L}\left(\omega\right)}{8}
e^{-2 \sigma_{\nu}
	^2-\frac{\Omega _3}{T}} \left(2 \sqrt{2}
\left(e^{\frac{2 \Omega _3}{T}}-1\right) \sin \left(2
\left(\theta _r\right)\right)+\right.
\end{equation}
\[
\left.\left(-4
e^{\frac{\Omega _3}{T}}+e^{\frac{2 \Omega
		_3}{T}}+1\right) \cos \left(2 \left(\theta
_r\right)\right)+e^{2 \sigma_{\nu} ^2} \left(4 e^{\frac{\Omega
		_3}{T}}+3 e^{\frac{2 \Omega
		_3}{T}}+3\right)\right)
\]
\small\begin{equation}
\langle\rho^{L=2}_{00}(\theta_r,\Omega_3,T)\rangle=\frac{\tilde{P}^{2}_{L}\left(\omega\right)}{40}
e^{-2 \sigma_{\nu}
	^2-\frac{\Omega _3}{T}} \left(2 \left(\sqrt{6}-6\right)
\left(e^{\frac{2 \Omega _3}{T}}-1\right) \sin \left(2
\left(\theta _r\right)\right)+\right.
\end{equation}
\[
\left.+2
e^{\frac{\Omega _3}{T}} \left(\left(2 \sqrt{6}-1\right)
\cosh \left(\frac{\Omega _3}{T}\right)+4\right) \cos
\left(2 \left(\theta
_r\right)\right)-\left(2 \sqrt{6}-13\right) e^{2 \sigma_{\nu}
	^2} \left(e^{\frac{2 \Omega _3}{T}}+1\right)+8 e^{2
	\sigma_{\nu} ^2+\frac{\Omega _3}{T}}\right)
\]
\item[The coefficient] $r_{10}$
\small\begin{equation}
\langle r^{L=0}_{10}(\phi_r,\theta_r,\Omega_3,T)\rangle=-\tilde{P}^{2}_{L}\left(\omega\right)\sqrt{2} e^{-2 \sigma_{\nu}
	^2}   \cos \left(\phi
_r\right) \sinh ^2\left(\frac{\Omega _3}{2 T}\right)
\sin \left(2 \left(\theta
_r\right)\right)
\end{equation}
\small\begin{equation}
\langle r^{L=1}_{10}(\phi_r,\theta_r,\Omega_3,T)\rangle=\frac{\tilde{P}^{2}_{L}\left(\omega\right)}{4} e^{-2 \sigma_{\nu}
	^2}   \cos \left(\phi
_r\right) \left(\sqrt{2} \left(\cosh \left(\frac{\Omega
	_3}{T}\right)-2\right) \sin \left(2
\left(\theta _r\right)\right)\right.
\end{equation}
\[
\left.-4 \sinh
\left(\frac{\Omega _3}{T}\right) \cos \left(2
\left(\theta
_r\right)\right)\right)
\]
\small\begin{equation}
\langle r^{L=2}_{10}(\phi_r,\theta_r,\Omega_3,T)\rangle=\frac{\tilde{P}^{2}_{L}\left(\omega\right)}{20} e^{-2
	\sigma_{\nu} ^2}   \cos
\left(\phi _r\right) \left(4 \sqrt{21-6 \sqrt{6}} \sinh
\left(\frac{\Omega _3}{T}\right) \cos \left(2
\left(\theta _r\right)\right)+\right.
\end{equation}
\[
\left.+\sqrt{2}
\left(\left(2 \sqrt{6}-1\right) \cosh
\left(\frac{\Omega _3}{T}\right)+4\right) \sin \left(2
\left(\theta
_r\right)\right)\right)
\]
\item[The coefficient] $\alpha_{10}$
\small\begin{equation}
\langle\alpha^{L=0}_{10}(\phi_r,\theta_r,\Omega_3,T)\rangle=-\frac{\tilde{P}^{2}_{L}\left(\omega\right)\sqrt{2}
	\sin \left(\phi
	_r\right) \sinh \left(\frac{\Omega _3}{T}\right) \sin
	\left(\theta _r\right)}{\sqrt{e^{\sigma_{\nu}
			^2}}}
\end{equation}
\small\begin{equation}
\langle\alpha^{L=1}_{10}(\phi_r,\theta_r,\Omega_3,T)\rangle=-\frac{ \tilde{P}^{2}_{L}\left(\omega\right)\sin \left(\phi _r\right)
	\left(\sqrt{2} \sinh \left(\frac{\Omega _3}{T}\right)
	\sin \left(\theta _r\right)+2 \cosh
	\left(\frac{\Omega _3}{T}\right) \cos
	\left(\theta _r\right)\right)}{2
	\sqrt{e^{\sigma_{\nu} ^2}}}
\end{equation}
\small\begin{equation}
\langle\alpha^{L=2}_{10}(\phi_r,\theta_r,\Omega_3,T)\rangle=-\frac{\tilde{P}^{2}_{L}\left(\omega\right)  \sin \left(\phi _r\right)}{10
	\sqrt{e^{\sigma_{\nu} ^2}}} \left(2
\left(\left(3 \sqrt{2}+\sqrt{3}\right) \cosh
\left(\frac{\Omega _3}{T}\right)+2 \sqrt{3}\right) \cos
\left(\theta _r\right)\right.
\end{equation}
\[
\left.-5 \sqrt{2} \sinh
\left(\frac{\Omega _3}{T}\right) \sin
\left(\theta _r\right)\right)
\]
\item[The coefficient] $r_{1,-1}$
\small\begin{equation}
\langle r^{L=0}_{1,-1}(\phi_r,\theta_r,\Omega_3,T)\rangle=\frac{\tilde{P}^{2}_{L}\left(\omega\right)}{2} e^{-2 \sigma_{\nu}
	^2}   \cos \left(2 \phi
_r\right) \sinh ^2\left(\frac{\Omega _3}{2 T}\right)
\left(e^{2 \sigma_{\nu} ^2}-\cos \left(2
\left(\theta
_r\right)\right)\right)
\end{equation}
\small\begin{equation}
\langle r^{L=1}_{1,-1}(\phi_r,\theta_r,\Omega_3,T)\rangle=-\frac{\tilde{P}^{2}_{L}\left(\omega\right)}{16}  \cos \left(2 \phi
_r\right) e^{-2 \sigma_{\nu} ^2-\frac{\Omega _3}{T}} \left(-2
\sqrt{2} \left(e^{\frac{2 \Omega _3}{T}}-1\right) \sin
\left(2 \left(\theta
_r\right)\right)\right.
\end{equation}
\[
\left.-\left(-4 e^{\frac{\Omega
		_3}{T}}+e^{\frac{2 \Omega _3}{T}}+1\right) \cos \left(2
\left(\theta _r\right)\right)+e^{2 \sigma_{\nu}
	^2} \left(4 e^{\frac{\Omega _3}{T}}+e^{\frac{2 \Omega
		_3}{T}}+1\right)\right)
\]
\small\begin{equation}
\langle r^{L=2}_{1,-1}(\phi_r,\theta_r,\Omega_3,T)\rangle=\frac{\tilde{P}^{2}_{L}\left(\omega\right)}{40} e^{-2
	\sigma_{\nu} ^2}  \cos
\left(2 \phi _r\right) \left(4 \left(\cos \left(2
\left(\theta _r\right)\right)+2 e^{2 \sigma_{\nu}
	^2}\right)+\right.
\end{equation}
\[
\left.+\cosh \left(\frac{\Omega _3}{T}\right)
\left(\left(2 \sqrt{6}-1\right) \cos \left(2
\left(\theta _r\right)\right)+\left(1+6
\sqrt{6}\right) e^{2 \sigma_{\nu} ^2}\right)+2
\left(\sqrt{6}-6\right) \sinh \left(\frac{\Omega
	_3}{T}\right) \sin \left(2 \left(\theta
_r\right)\right)\right)
\]
\item[The coefficient] $\alpha_{1,-1}$
\small\begin{equation}
\langle\alpha^{L=0}_{1,-1}(\phi_r,\theta_r,\Omega_3,T)\rangle=\frac{\tilde{P}^{2}_{L}\left(\omega\right)}{2} e^{-2 \sigma_{\nu}
	^2}  \sin \left(2 \phi
_r\right) \sinh ^2\left(\frac{\Omega _3}{2 T}\right)
\left(\cos \left(2 \left(\theta
_r\right)\right)-e^{2 \sigma_{\nu} ^2}\right)
\end{equation}
\small\begin{equation}
\langle\alpha^{L=1}_{1,-1}(\phi_r,\theta_r,\Omega_3,T)\rangle=\frac{\tilde{P}^{2}_{L}\left(\omega\right)}{16}
\sin \left(2 \phi
_r\right) e^{-2 \sigma_{\nu} ^2-\frac{\Omega _3}{T}} \left(-2
\sqrt{2} \left(e^{\frac{2 \Omega _3}{T}}-1\right) \sin
\left(2 \left(\theta
_r\right)\right)\right.
\end{equation}
\[
\left.-\left(-4 e^{\frac{\Omega
		_3}{T}}+e^{\frac{2 \Omega _3}{T}}+1\right) \cos \left(2
\left(\theta _r\right)\right)+e^{2 \sigma_{\nu}
	^2} \left(4 e^{\frac{\Omega _3}{T}}+e^{\frac{2 \Omega
		_3}{T}}+1\right)\right)
\]
\small\begin{equation}
\langle\alpha^{L=2}_{1,-1}(\phi_r,\theta_r,\Omega_3,T)\rangle=-\frac{\tilde{P}^{2}_{L}\left(\omega\right)}{40} e^{-2
	\sigma_{\nu} ^2}  \sin
\left(2 \phi _r\right) \left(4 \left(\cos \left(2
\left(\theta _r\right)\right)+2 e^{2 \sigma_{\nu}
	^2}\right)+\right.
\end{equation}
\[
\left.+\cosh \left(\frac{\Omega _3}{T}\right)
\left(\left(2 \sqrt{6}-1\right) \cos \left(2
\left(\theta _r\right)\right)+\left(1+6
\sqrt{6}\right) e^{2 \sigma_{\nu} ^2}\right)+2
\left(\sqrt{6}-6\right) \sinh \left(\frac{\Omega
	_3}{T}\right) \sin \left(2 \left(\theta
_r\right)\right)\right)
\]
\end{description}
The figures in section \ref{spincoal} are based on these formulae.
Note that, as a consistency check, Eq. \ref{phirdef} is recovered from the last two equations for all studied case of $L$.

\end{document}